\documentclass[runningheads]{llncs}
\usepackage[T1]{fontenc}
\usepackage{graphicx}
\usepackage{cite}
\usepackage{color, soul}
\usepackage{xcolor,url}
\usepackage{subfigure}
\usepackage{multirow}
\graphicspath{ {./images/} }
\usepackage{flushend} 
\usepackage{ulem}    
\usepackage{longtable}
\usepackage{xparse}
\usepackage{booktabs}
\usepackage{subfigure}
\usepackage{caption}
\usepackage{array}
\usepackage{diagbox}
\usepackage{color}

\begin{document}
\title{MnTTS2: An Open-Source Multi-Speaker Mongolian Text-to-Speech Synthesis Dataset}
\titlerunning{MnTTS2: Multi-Speaker Mongolian Text-to-Speech Synthesis Dataset}
%
\author{Kailin Liang$^{\dag}$ \and
Bin Liu$^{\dag}$ \and Yifan Hu$^{\dag}$ \and Rui Liu\thanks{: Corresponding Author is Rui Liu. $^{\dag}$: Equal Contributions. This research was funded by the High-level Talents Introduction Project of Inner Mongolia University (No. 10000-22311201/002) and the Young Scientists Fund of the National Natural Science Foundation of China (No. 62206136).} \and Feilong Bao \and
Guanglai Gao }
\authorrunning{Kailin Liang \& Bin Liu \& Yifan Hu et al.}
%
\institute{Inner Mongolia University, Hohhot, China\\
\email{liangkailin98@foxmail.com,iframe\_liu@163.com,hyfwalker@163.com}
\email{liurui\_imu@163.com, \{csfeilong, csggl\}@imu.edu.cn}}

\maketitle              

\begin{abstract}
Text-to-Speech (TTS) synthesis for low-resource languages is an attractive research issue in academia and industry nowadays. 
Mongolian is the official language of the Inner Mongolia Autonomous Region and a representative low-resource language spoken by over 10 million people worldwide.
However, there is a relative lack of open-source datasets for Mongolian TTS.
Therefore, we make public an open-source multi-speaker Mongolian TTS dataset, named MnTTS2, for the benefit of related researchers. 
In this work, we prepare the transcription from various topics and invite three professional Mongolian announcers to form a three-speaker TTS dataset, in which each announcer records 10 hours of speeches in Mongolian, resulting 30 hours in total. Furthermore, we build the baseline system based on the state-of-the-art FastSpeech2 model and HiFi-GAN vocoder. The experimental results suggest that the constructed MnTTS2 dataset is sufficient to build robust multi-speaker TTS models for real-world applications. The MnTTS2 dataset, training recipe, and pretrained models are released at: \url{https://github.com/ssmlkl/MnTTS2}.


\end{abstract}

\section{Introduction}
Text-to-Speech (TTS) aims to convert the input text to human-like speech \cite{shen2018natural}. It is a standard technology in human-computer interaction, such as cell phone voice assistants, car navigation, smart speakers, etc. 
The field of speech synthesis has developed rapidly in recent years.
Different from the traditional methods, which use concatenation \cite{1168657}, statistical modeling \cite{rabiner1989tutorial} based methods to synthesize speech, neural end-to-end TTS models achieve remarkable performance with the help of Encoder-Decoder architecture \cite{cho2014properties}. Typical models include Tacotron~ \cite{wang2017tacotron}, Tacotron2~ \cite{shen2018natural}, Transformer TTS~ \cite{li2019neural}, Deep Voice~ \cite{arik2017deep}, etc. To further accelerate the inference speed, the non-autoregressive TTS models, such as FastSpeech~ \cite{ren2019fastspeech}, FastSpech2(s)~ \cite{ren2020fastspeech}, are proposed and become the mainstream methods of TTS. Note that armed with the neural network based vocoder, including WaveNet~ \cite{oord2016wavenet}, WaveRNN~ \cite{kalchbrenner2018efficient}, MelGAN~ \cite{kumar2019melgan}, HiFi-GAN~ \cite{kong2020hifi}, etc., the TTS model can synthesize speech sounds that are comparable to human sounds.

We note that an important factor in the rapid development of neural TTS mentioned above is the large scale corpus resources.
This is especially true for languages such as English and Mandarin, which are widely spoken worldwide. However, low-resource language such as Mongolian \cite{bulag2003mongolian} have been making slow progress in related research due to the difficulties in corpus collection. Therefore, building a large-scale and high-quality Mongolian TTS dataset is necessary. In addition, our lab have previously open-sourced a single-speaker dataset called MnTTS \cite{hu2022mntts}, which was recorded by a young female native Mongolian speaker and received much attention from academia and industry upon its release. This also shows the necessity of continuing to collect and organize Mongolian speech synthesis datasets and opening the baseline model's source code.

Motivated by this, this paper presents a multi-speaker Mongolian TTS dataset, termed as  MnTTS2, which increases the number of speakers to three and increases the data size from 8 to 30 hours, with an average of 10 hours per speaker. The textual content has been further expanded and enriched in the domain. Similar with our MnTTS, MnTTS2 dataset is freely available to academics and industry practitioners.

To demonstrate the reliability of MnTTS2, we combined the state-of-the-art FastSpeech2 \cite{ren2020fastspeech} model and the HiFi-GAN \cite{kong2020hifi} vocoder to build the accompanied baseline model for MnTTS2. We conduct listening experiments and report the Naturalness Mean Opinion Score (N-MOS) and Speaker Similarity Mean Opinion Score (SS-MOS) results in terms of naturalness and speaker similarity respectively.
The experimental results show that our system can achieve satisfactory performance on the MnTTS2, which indicates that 
the MnTTS2 corpus is practically usable and can be used to build robust multi-speaker TTS system.

The main contributions are summarized as follows.
1) We developed a multi-speaker TTS dataset, termd as MnTTS2, containing three speakers. The total audio duration is about 30 hours. The transcribed text covers various domains, such as sports and culture, etc.
2) We used the state-of-the-art non-autoregressive FastSpeech2 model to build the baseline model and validate our MnTTS2.
3) The MnTTS2 dataset, source code, and pre-trained models will be publicly available to academics and industry practitioners.

The rest of the paper is organized as follows. Section 2 revisits the related works about the Mongolian TTS corpus. In Section 3, we introduce the details of MnTTS2, including the corpus structure and statistical information. Section 4 explains and discusses the experimental setup and experimental results. Section 5 discusses the challenges faced by Mongolian speech synthesis and the future research directions. Section 6 concludes the paper and summarizes the work and research of this paper.

\section{Related Work}

For mainstream languages such as English and Mandarin, there are many free and publicly available TTS datasets.
For example, LJSpeech \cite{ito2017lj} is a single-speaker dataset for English.
To rich the speaker diversity, some multi-speaker TTS dataset are released, such as  Libritts \cite{zen2019libritts} for English and Aishell \cite{https://doi.org/10.48550/arxiv.2010.11567} for Chinese. 

For the low-resource language such as Mongolian, the available resources are pretty limited. We note that some attempts tried to improve the effect of TTS synthesis under low resource data with unsupervised learning \cite{barlow1989unsupervised}, semi-supervised learning \cite{zhu2005semi}, and transfer learning \cite{weiss2016survey} methods, etc. However, due to the lack of large-scale training data, all the mentioned methods are difficult to achieve the effect that meets the requirements of practical scenarios. 

In order to promote the development of Mongolian TTS, some works built their own Mongolian TTS corpus and designed various models to achieve good results. For example, 
Huang et al. established the first emotionally controllable Mongolian TTS system and achieved eight emotional embeddings by transfer learning and emotional embedding \cite{huang2021mongolian}.
Rui Liu et al. introduced a new method to segment Mongolian words into stems and suffixes, which greatly improved the performance of the Mongolian rhyming phrase prediction system \cite{liu2016mongolian}.
Immediately after that, Rui Liu proposed a DNN-based Mongolian speech synthesis system, which performs better than the traditional HMM \cite{liu2017mongolian}. Also, he introduced the Bidirectional Long Term Memory (BiLSTM) model to improve the phrase break prediction step in the traditional speech synthesis system, making it more applicable to Mongolian \cite{liu2018improving}.
Unfortunately, none of the Mongolian TTS dataset from the above works have been released publicly and are not directly available to the public. We also found that some dataset in related fields, such as M2ASR-MONGO \cite{DBLP:conf/ococosda/ZhiSDLW21} for Mongolian speech recognition, are public recently.
However, the speech recognition corpus cannot be applied in the TTS filed due to the environment noise and improper speaking style issues etc.

We previously released the single-speaker MnTTS dataset \cite{hu2022mntts}, called MnTTS. The total duration of the MnTTS is 8 hours, and it was recorded in a studio by a professional female native Mongolian announcer. However, the duration and speaker diversity still needs to be further expanded.
In a nutshell, it is necessary to construct a high-quality multi-speaker Mongolian TTS dataset to further promote the Mongolian TTS research, which is the focus of this paper.
We will introduce the details of the MnTTS2 at the following subsection.

\section{MnTTS2 Dataset}
In this section, we first revisit the MnTTS dataset briefly and then introduce our MnTTS2 by highlighting the extended content.  

\subsection{MnTTS}
In the preliminary work, we presented a high-quality single-speaker Mongolian TTS dataset, called MnTTS \cite{hu2022mntts}. The transcription of the dataset was collected from a wide range of topics, such as policy, sports, culture, etc. The Mongolian script was then converted to Latin sequences to avoid as many miscoding issues as possible. A professional female native Mongolian announcer was invited to record all the audio. A Mongolian volunteer was invited to check and re-align the alignment errors. The audio containing ambient noise and mispronunciation was removed to ensure the overall quality.

MnTTS has received much attention from researchers in the same industry upon its release. Furthermore, the subset was used in the Mongolian Text-to-Speech Challenge under Low-Resource Scenario at NCMMSC2022 \footnote{\url{http://mglip.com/challenge/NCMMSC2022-MTTSC/index.html}}. The organizers provided two hours of data for all participants to train their models. This competition also promotes the development of intelligent information processing in minority languages within China.

\subsection{MnTTS2}
The construction pipeline of MnTTS2 consists of ``Text collection and narration'', ``Text preprocessing'' and  ``Audio recording and audio-text alignment''. We will introduce them in order and then report the corpus structure and statistics.

\subsubsection{Text collection and narration}

Similar with MnTTS \cite{hu2022mntts}, the first step in building the MnTTS2 dataset is to collect a large amount of transcription. The natural idea for collecting such a text materials is crawl text information from websites and electronic books. The text topics should cover human daily usage scenarios as much as possible. Following this, we crawled 23, 801 sentences, that are rich in content and have a wide range of topics (e.g., politics, culture, economy, sports, etc.), to meet our requirements well. At the same time, we manually filtered and removed some texts with unsuitable content, which may involve sensitive political issues, religious issues or pornographic content. These contents are removed in the hope that our dataset can make a positive contribution to the development of Mongolian language, which is the original intention of our work.

\subsubsection{Text preprocessing}
Compared to mainstream languages, such as Mandarin and English, traditional Mongolian performs agglutinative characteristic \cite{6639250}. This makes the Mongolian letters express different styles in different contexts and brings a serious harmonic phenomenon \cite{hu2022mntts}. In order to solve this problem, we transformed the texts into a Latin alphabet, instead of traditional Mongolian representation, for TTS training.
The entire pipeline of converting Mongolian texts into Latin sequences is divided into three steps: encoding correction, Latin conversion and text regularization. The detailed description can be found in our previous work MnTTS \cite{hu2022mntts}.

\begin{figure}[h!]
\centering
\includegraphics[width=\linewidth]{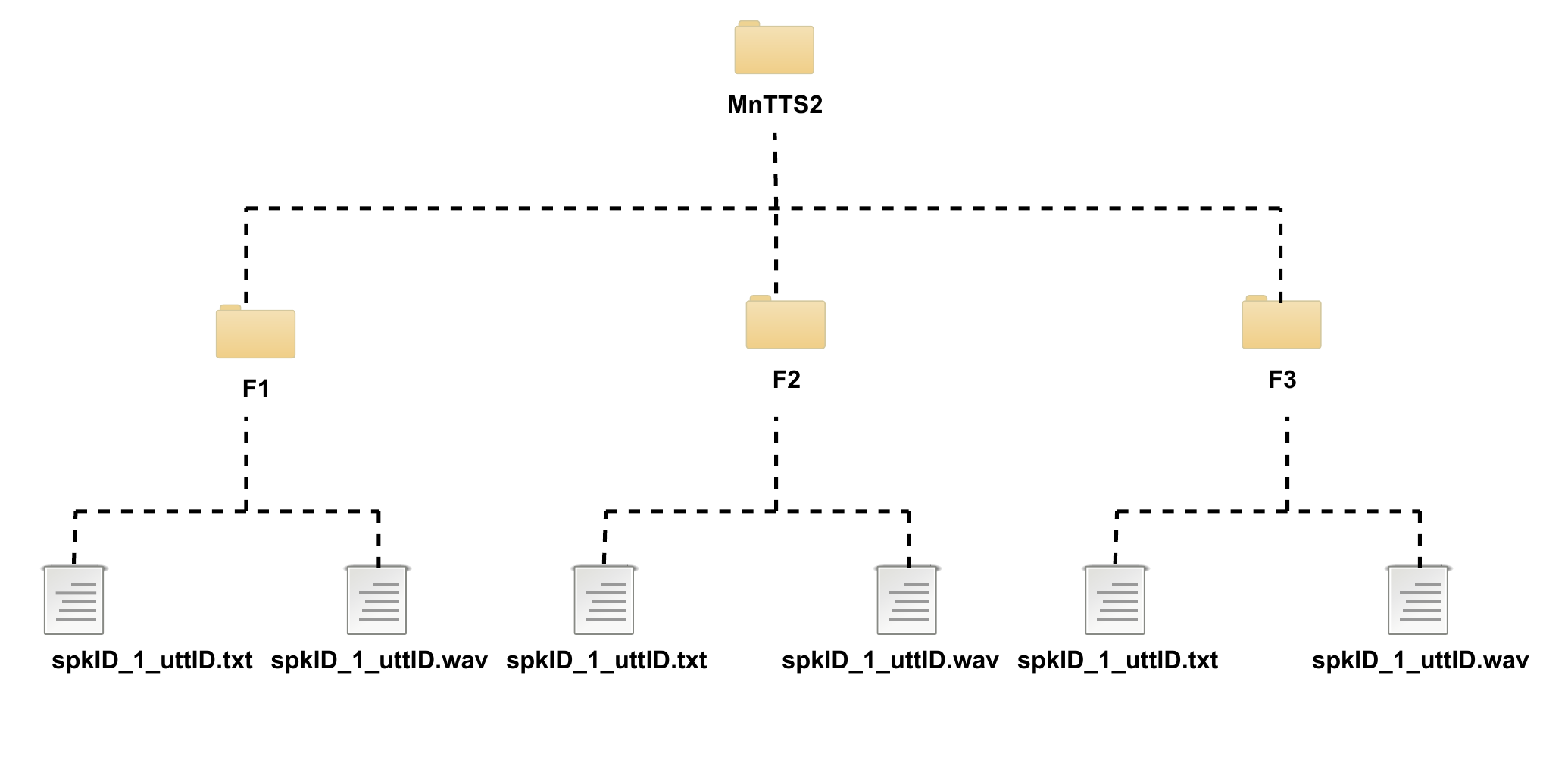}
\caption{The folder structure of the MnTTS2 corpus.}
\label{figure1}
\end{figure}

\subsubsection{Audio recording and audio-text alignment}
Different with the MnTTS \cite{hu2022mntts}, we invited three native Mongolian-speaking announcers to record the audio. Each announcer volunteered to participate and signed an informed consent form to be informed of the data collection and use protocol. F1, F2 and F3 are three native Mongolian speaking females, with F2 being a little girl and F1 and F3 being slightly older in grade.
All recordings were done in a standard recording studio at Inner Mongolia University. We choose Adobe Audition \footnote{\url{https://www.adobe.com/cn/products/audition.html}} as the recording software.

During the recording process, we asked the announcer to keeps a 0.3s pause at the beginning and end of each audio segment, keeps a constant distance between the lips and the microphone, performs a slight pause at the comma position, and performs an appropriate pitch boost at the question mark position.

To ensure the quality of the recording data , we rechecked the recording data after completing the recording work. Specifically, we invited three volunteers to check each text against its corresponding natural audio. 
These volunteers are responsible for splitting the recorded audio file into sentences and aligning the split sentences with the text. The Mongolian text is represented by a Latin sequence, where each Latin word in the sequence becomes a word and each letter that makes up the word is called a character. Characters also include punctuation marks, such as commas (`,'), periods (`.') , question mark (`?') , exclamation mark (`!') etc. Finally, we obtained about 30h of speech data, which were sampled at 44.1kHz with a sampling accuracy of 16bit.

\begin{table}[h!]
\setlength{\abovecaptionskip}{10pt} 
\setlength{\belowcaptionskip}{-10pt}
\caption{The statistics results of MnTTS2 dataset.}
\label{table1}
\centering
\begin{tabular}{p{30mm}<{\centering} p{10mm}<{\centering} p{15mm}<{\centering} p{15mm}<{\centering} p{15mm}<{\centering} p{15mm}<{\centering}}
\toprule
\diagbox[]{Statistical Unit}{Speaker ID} &&    & F1 & F2 & F3 \\ \hline
\multirow{4}{*}{Character} &&  Total & 572016 & 459213 & 601366 \\
 &&  Mean & 79 & 61 & 67 \\
 &&  Min & 12 & 2 & 2 \\
 &&  Max & 189 & 188 & 190 \\ \hline
\multirow{4}{*}{Word} &&  Total & 88209 & 71245 & 92719 \\
 && Mean & 12 & 9 & 10 \\
 && Min & 3 & 1 & 1 \\
 && Max & 29 & 30 & 29 \\ \bottomrule
\end{tabular}
\end{table}

\begin{figure}[h!]
 \centering

\begin{minipage}{\linewidth}
  \centerline{
  \includegraphics[width=.32\linewidth]{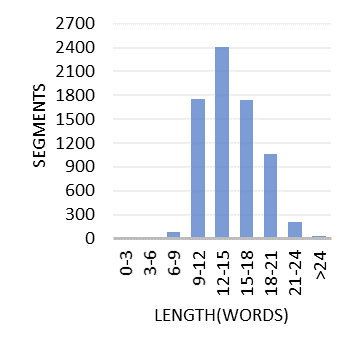}
  \includegraphics[width=.32\linewidth]{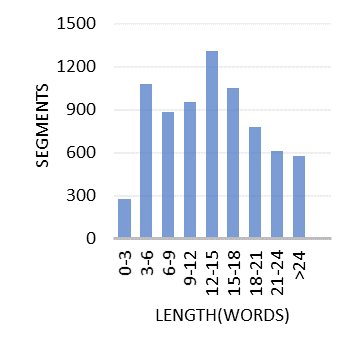}
  \includegraphics[width=.32\linewidth]{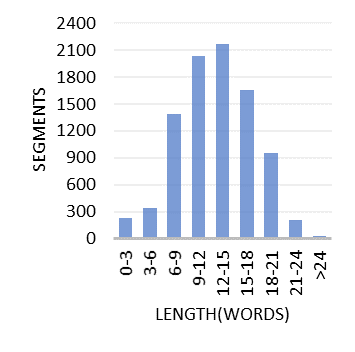}
  }
 \qquad \qquad \qquad \qquad (a) F1  \qquad \qquad \qquad\qquad(b) F2  \qquad\qquad\qquad \qquad (c) F3 \\
\end{minipage}
\hfill

\begin{minipage}{\linewidth}
  \centerline{
  \includegraphics[width=.32\linewidth]{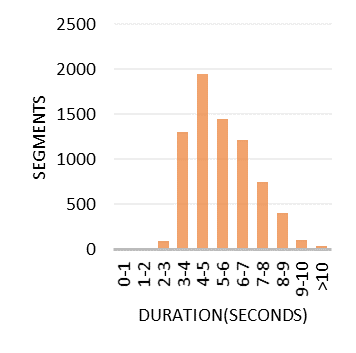}
  \includegraphics[width=.32\linewidth]{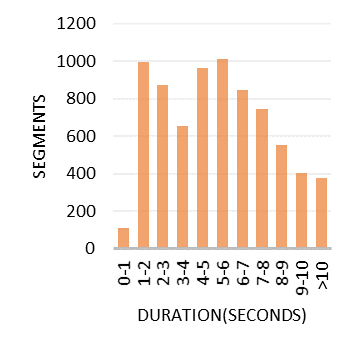}
  \includegraphics[width=.32\linewidth]{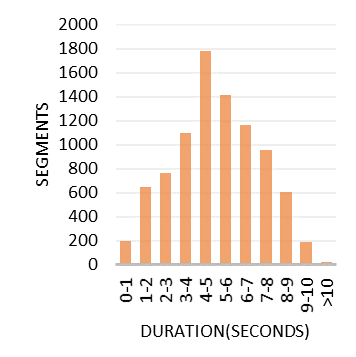}
  }
 \qquad \qquad \qquad \qquad (a) F1  \qquad \qquad \qquad\qquad(b) F2 \qquad\qquad\qquad \qquad (c) F3 \\
  \vspace{-3mm}
\end{minipage}
\vfill
 \caption{Word number distributions  (a, b, c) and sentence duration distributions   (d, e, f) for all speakers of MnTTS2.}
        \label{figure2}
        \vspace{-3mm}
\end{figure}

\subsubsection{Corpus structure and statistics}
The file structure of the MnTTS2 corpus is shown in Fig. \ref{figure1}. Each speaker's recording file and the corresponding text collection are saved in a folder named after the speaker. All audios are stored in WAV format files , sampled at 44.10kHz, and coded in 16 bits.
All text is saved in a TXT file encoded in UTF-8. The file name of the audio is the same as the corresponding text file name, and the name of each file consists of the speaker, document ID, and corpus ID.

The statistical results of the MnTTS2 data are shown in Table \ref{table1} and Fig. \ref{figure2}. As shown in Table \ref{table1}, the entire corpus has a total of 23, 801 sentences. Take the speaker F1 for example, F1 with a total of 572,016 Mongolian characters, and the average number of characters per sentence is 79, with the shortest sentence having 12 characters and the longest sentence having 189 characters. If words are used as the statistical unit, the total number of words in this dataset for F1 is 88,209, the mean value of words in each sentence is 12, the minimum value is 3, and the maximum value is 29. As shown in Fig. \ref{figure2}, we also counted the sentence duration to draw a histogram. Take speaker F1 for example, the word numbers of the sentences are concentrated in 12-15, and duration are concentrated in 4-5 seconds. In comparison, we found that the word numbers of sentences for F2 was not particularly concentrated, and the duration were relatively scattered. F3, on the other hand, is more similar to F1, with a more obvious concentration. The statistics of all three speakers are in line with the normal distribution.

\section{Speech Synthesis Experiments}
To verify the validity of our MnTTS2, we conducted Mongolian TTS experiments based on the FastSpeech2 model and HiFi-GAN vocoder and evaluated the synthesized speech using Mean Opinion Score (MOS) metric in terms of naturalness and speaker similarity.

\subsection{Experimental Setup}

\begin{figure}[h!]
\centering
\includegraphics[width=\linewidth]{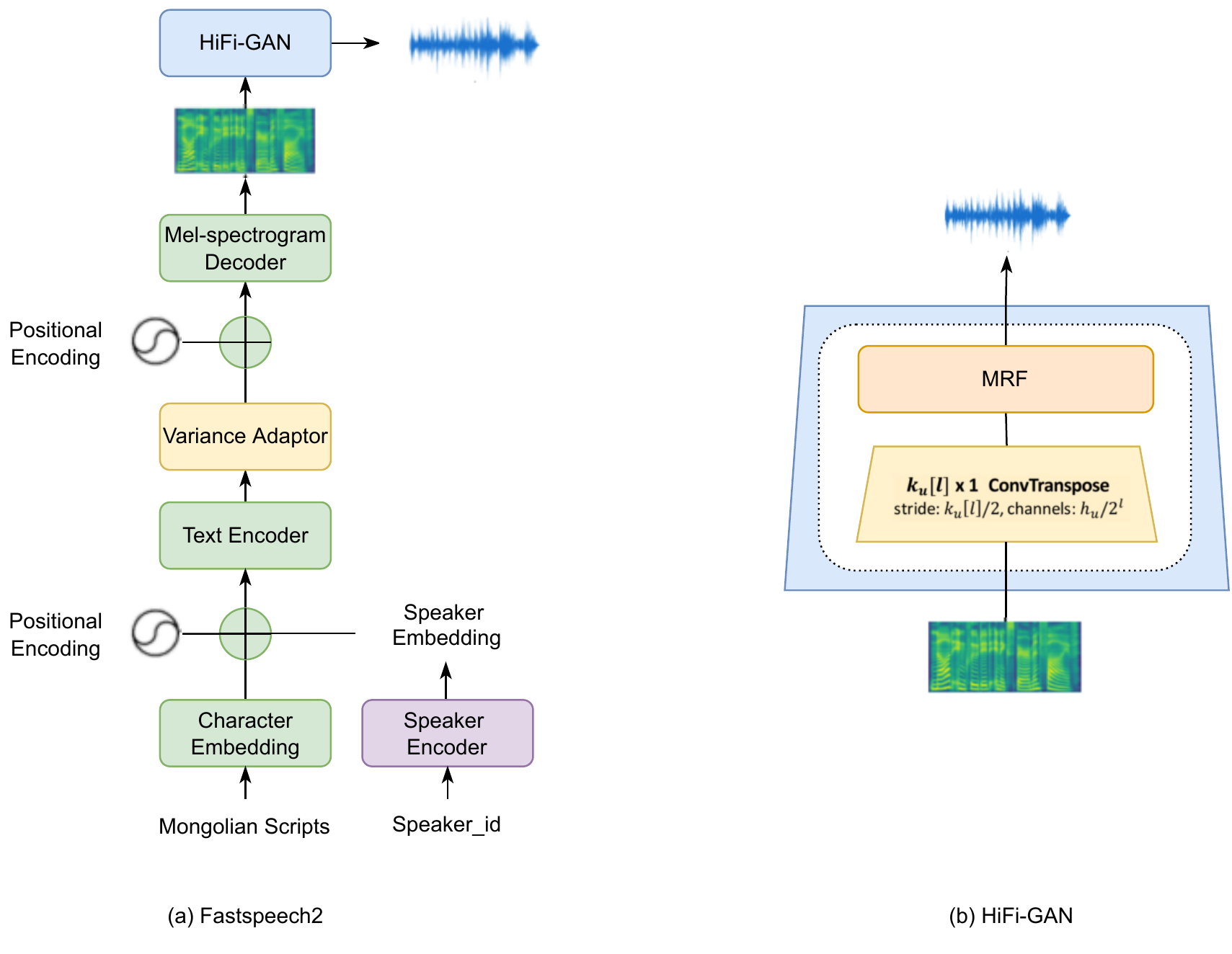}
\caption{The structure of the FastSpeech2+HiFi-GAN model. We implement the multi-speaker FastSpeech2 by adding the speaker encoder module.}
\label{model}
\end{figure}

We use the TensorFlowTTS toolkit \footnote{\label{ttscode}\url{https://github.com/TensorSpeech/TensorFlowTTS}} to build an end-to-end TTS model based on the FastSpeech2 model. The FastSpeech2 model converts the input Mongolian text into Mel-spectrogram features, and then the HiFi-GAN vocoder reconstructs the waveform by the Mel-spectrogram features. FastSpeech2 is a state-of-the-art non-autoregressive \cite{gu2017non} speech synthesis model that extracts duration, pitch, and energy directly from the speech waveform and uses these features as input conditions in training.  This model can effectively solve errors such as repetition and word skipping, and has the advantage of fast training speed. FastSpeech2 introduces more variance information to alleviate the one-to-many mapping problem. Also, the pitch prediction is improved by wavelet transform. Most of all, FastSpeech2 has the characteristics of fast, robust, and controllable speech synthesis. This is the main reason why we choose FastSpeech2. As shown in Fig. \ref{model}, based on FastSpeech2, we implement the multi-speaker FastSpeech2 by adding the speaker encoder module. The speaker encoder includes speaker embedding, dense, and softplus layers $^3$.
In the network architecture setting, the number of speakers is 3. The dimension of speaker embedding is 384. The hidden side of the text encoder is 384 and the number of hidden layers is 4, the hidden layer size of the decoder is 384 and the number of hidden layers is 4. The number of Conv layers of the variance predictors is 2 and the dropout rate is 0.5. The initial learning rate is 0.001 and the dropout rate is 0.2.

The HiFi-GAN vocoder builds the network through a generative adversarial network to converts Mel-spectrogram into high-quality audio. The generator of HiFi-GAN consists of an upsampling structure, which consists of a one-dimensional transposed convolution, and a multi-receptive filed fusion module, which is responsible for optimizing the upsampling points. HiFi-GAN, as a generative adversarial network, has two kinds of discriminators, including multi-scale and multi-period discriminators. The generagor kernel size of HiFi-GAN is 7 and the upsampling ratio is (8,8,2,2). The list of discriminators for the cycle scale is (2,3,5,7,11). The Conv filters of each periodic discriminator are 8. The pooling type of output downsampling in the melgan discriminator is AveragePooling1D, the kernel size is (5,3), and the activation function is LeakyReLU. HiFi-GAN is trained independently of FastSpeech2. For each speaker, the generator with only stft loss is first trained for 100k steps, and then the generator and discriminator are trained for 100k steps. This gives us the corresponding vocoder for each of the three speakers.

Note that a teacher Tacotron2 model trained for each speaker was used to extract duration from the attention contrast for subsequent FastSpeech2 model training. For each speaker, the Tacotron2 model trained with 100k steps of MnTTS was used to extract the duration. After that, the multi-speaker FastSpeech2 model was trained with 200k steps to do the final speech generation. 100k steps were trained for HiFi-GAN's generator and 100k steps for jointly training the generator and discriminator. All the above models were trained on 2 Tesla V100 GPUs.

\subsection{Naturalness Evaluation}
For a full comparison of naturalness, we compared our baseline system, \textbf{FastSpeech2+HiFi-GAN} with the \textbf{Ground Truth} speech. In addition, to verify the performance of HiFi-GAN, we added a \textbf{FastSpeech2+Griffin-Lim} baseline model for further comparison. The Griffin-Lim algorithm can directly obtain the phase information of the audio to reconstruct the waveform without additional training. 
We used the Naturalness Mean Opinion Score (SS-MOS) \cite{streijl2016mean} to assess naturalness. For each speaker, we randomly select 20 sentences as the evaluation set, which are not used for training. The model-generated audio and the ground truth audio were randomly disrupted and distributed to listeners. During the evaluation process, 10 native Mongolian speakers were asked to evaluate the naturalness of the generated 400 audio speeches in a quiet environment. 

The N-MOS results are given in Table \ref{tab2}. Ground truth speech gets the best performance without a doubt. FastSpeech2+HiFi-GAN outperforms the FastSpeech2+Griffin-Lim and achieves much closer performance to the ground truth. 
Each speaker's N-MOS score was above 4.0 on the combination of FastSpeech2 and HiFi-GAN.

Specifically, for the F1, F2, and F3, the N-MOS of FastSpeech2+HiFi-GAN achieved 4.02, 4.15, and 4.29 respectively, which is encouraging. This proves that high-quality Mongolian speech can be synthesized using MnTTS2 and the proposed model. 
In a nutshell, all results prove that our MnTTS2 dataset can be used to build a robust TTS system for high-quality speech generation.

\begin{table}[h]
\setlength{\abovecaptionskip}{10pt} 
\setlength{\belowcaptionskip}{-10pt}
\centering
\caption{\label{tab2}Naturalness mean opinion score (N-MOS) results for all systems with 95$\%$ Confidence intervals.}
\begin{tabular}{p{43mm}<{\centering} p{25mm}<{\centering} p{25mm}<{\centering} p{25mm}<{\centering} p{25mm}<{\centering}}
\toprule
\diagbox[]{System}{Speaker ID} & F1 & F2 & F3 \\ \hline
FastSpeech2+Griffin-Lim  &  3.56$\pm$0.18  &   3.59$\pm$0.04     &   3.86$\pm$0.12      \\
\textbf{FastSpeech2+HiFi-GAN}    & \textbf{4.02$\pm$0.18}   &\textbf{ 4.15$\pm$0.06 }      & \textbf{4.29$\pm$0.11}        \\ \hline
Ground Truth             & \textbf{4.73$\pm$0.08}   & \textbf{4.70$\pm$0.14}       & \textbf{4.68$\pm$0.09 }         \\  \bottomrule
\end{tabular}

\end{table}

\subsection{Speaker Similarity Evaluation}
\textcolor{black}{We further conduct listening experiments to evaluate the speaker similarity performance for the FastSpeech+HiFi-GAN baseline system. The Speaker Similarity Mean Opinion Score (SS-MOS) results are reported in Table \ref{tab3}}.

We synthesized 20 audios for each speaker by FastSpeech2+HiFi-GAN baseline system. Ten native Mongolian-speaking volunteers were also invited to participate in the scoring. Each volunteer needs to evaluate whether the speaker is the same person or not in the synthesized and the ground truth audio.
The SS-MOS scores for F1, F2, and F3 are 4.58, 4.04, and 4.12 respectively, which is encouraging. The results show that the audio synthesized by the FastSpeech2+HiFi-GAN system performs good performance in terms of speaker similarity. The highest SS-MOS score was obtained for speaker F1. Auditioning the audio, we can find that F1's timbre has significant characteristics and the synthesized audio represents the speaker's voice information well. 
In a nutshell, this experiment shows that the MnTTS2 dataset can be used for speech synthesis work in multi-speaker scenarios.

\begin{table}[h]
\setlength{\abovecaptionskip}{10pt} 
\setlength{\belowcaptionskip}{-10pt}
\centering
\caption{\label{tab3}Speaker Similarity Mean opinion score (SS-MOS) results for FastSpeech2+HiFi-GAN system with 95$\%$ Confidence intervals.}
\begin{tabular}{p{40mm}<{\centering} p{25mm}<{\centering} p{25mm}<{\centering} p{25mm}<{\centering} p{25mm}<{\centering}}
\toprule
\diagbox[]{System}{Speaker ID} & F1 & F2 & F3 \\ \hline
FastSpeech2+HiFi-GAN   &  4.58$\pm$0.21  &   4.04$\pm$0.16     &   4.12$\pm$0.10      \\
    \bottomrule
\end{tabular}
\end{table}

\section{Challenges and Future Work}

With the development of ``Empathy AI'', the research of emotional TTS has attracted more and more attention \cite{9767637}. Speech synthesis in conversational scenarios and emotional speech synthesis are hot research topics nowadays \cite{liu2021reinforcement}. Furthermore, how to control the
emotion category and the emotional intensity during speech generation is an interesting direction \cite{https://doi.org/10.48550/arxiv.2210.15364}. However, our MnTTS2 does not involve information related to
emotion category and emotion intensity. In future work, we will carry out a more
comprehensive and in-depth expansion of the data to serve the development of
emotional Mongolian TTS.

\section{Conclusion}
We presented a large-scale, open-source Mongolian text-to-speech corpus, MnTTS2, which enriches MnTTS with more durations, topics, and speakers. Releasing our corpus under the Knowledge Attribution 4.0 International License, the corpus allows both academic and commercial use. We describe in detail the process of building the corpus, while we validate the usability of the corpus by synthesizing sounds with the FastSpeech2 model and the HiFi-GAN vocoder. 
The experimental results show that our system can achieve satisfactory performance on the MnTTS2, which indicates that the MnTTS2 corpus is practically usable and can be used to build robust multi-speaker TTS system.
In future work, we will introduce emotional TTS dataset to further enrich our corpus. We also plan to compare the effects of different TTS architectures and model hyperparameters on the results and conduct subsequent analyses.


%
\normalem 
\bibliographystyle{unsrt}
\bibliographystyle{splncs04}
\bibliography{ref_unsort}
%
\end{document}